\documentstyle[twocolumn,prl,aps,graphicx]{revtex}
\newcommand{\be}{\begin{eqnarray}}
\newcommand{\ee}{\end{eqnarray}}

\begin{document}
\draft \twocolumn[\hsize\textwidth\columnwidth\hsize\csname
@twocolumnfalse\endcsname
\title{A Hierarchically-Organized Phase Diagram near a Quantum Critical Point in
URu$_{2}$Si$_{2}$}
\author{K. H. Kim$^{1,*}$, N. Harrison$^{1}$, M. Jaime$^{1}$, G. S. Boebinger$^{1}$, and J. A. Mydosh$^{2,3}$}
\address{$^{1}$National High Magnetic Field Laboratory, MS-E536 LANL, Los Alamos, NM 87545.}
\address{$^{2}$Kamerlingh Onnes Laboratory, Leiden University, NL-2300 RA, Leiden, The Netherlands}
\address{$^{3}$Max-Planck Institut for Chemical Physics of Solids, N\"{o}thnitzer St. 40, D-01187 Dresden, Germany}
\maketitle

\begin{abstract}
A comprehensive transport study, as a function of both temperature
and magnetic field in continuous magnetic fields up to 45 T
reveals that URu$_{2}$Si$_{2}$ possesses all the essential
hallmarks of quantum criticality at temperatures above 5.5 K and
fields around 38 T, but then collapses into multiple low
temperature phases in a hierarchically-organized phase diagram as
the temperature is reduced.  Although certain generic features of
the phase diagram are very similar to those in the cuprates and
heavy fermion superconductors, the existence of multiple ordered
hysteretic phases near the field-tuned quantum critical point is
presently unique to URu$_{2}$Si$_{2}$. This finding suggests the
existence of many competing order parameters separated by small
energy difference in URu$_{2}$Si$_{2}$.

\end{abstract}

 \vskip0.1pc]
A common picture emerging in strongly correlated metals is that
exotic superconductivity may have as much to do with quantum
criticality \cite{laughlin01,hertz76,millis93} as with
unconventional pairing mechanisms \cite{mathur98,saxena00}. In our
modern understanding of phase transitions, \emph{criticality}
refers to the characteristic (often power-law) temperature
dependences of physical properties that occur in the vicinity of
phase transitions and arise from thermal fluctuations.
\emph{Quantum criticality} refers to our emerging understanding of
phase transitions that occur at zero temperature (quantum critical
points) in which quantum fluctuations play an important role. The
abundance of low energy excitations that persist near quantum
critical points provide a perfect recipe for instability, even in
the limit of zero temperature. This provides extra opportunities
to form new phases that would otherwise not exist.

Quantum criticality can be explicitly produced by depressing a
phase transition toward absolute zero, by tuning an external
control parameter, such as hydrostatic pressure \cite{mathur98} or
chemical composition \cite{lohneysen94} or, as in our experiments,
magnetic field. The physical properties of the system can then be
dominated by the critical fluctuations associated with this
quantum phase transition, or quantum critical point (QCP), even at
very high temperatures up to and above room temperature
\cite{sachdev99}. Furthermore, proximity to a QCP is known to give
rise to novel ground states, such as magnetism \cite{pfleiderer97}
and unconventional forms of superconductivity
\cite{mathur98,saxena00}. It might also explain high temperature
superconductivity and non-Fermi liquid behaviour observed in the
cuprates \cite{tallon01}. Since quantum criticality is becoming
increasingly recognized as a universal phenomenon in condensed
matter physics, it is of paramount importance to categorize and
understand how QCPs can stabilize new phases \cite{kivelson01}.

Recently, a magnetic field-tuned QCP was identified in the
ruthenate metal Sr$_{3}$Ru$_{2}$O$_{7}$ as an anomalous
resistivity deviating from the standard quadratic
temperature-dependence seen in conventional metals
\cite{grigera01}. While this work establishes magnetic fields
($H$) as a powerful tuning parameter for revealing QCPs in
correlated metals, a low-temperature ordered phase was not
identified near the QCP of Sr$_{3}$Ru$_{2}$O$_{7}$. The discovery
of an example of a magnetic-field-tuned QCP that actually gives
rise to new ordered phases would help firmly establish quantum
fluctuations as a critical factor in low-temperature phase
formation.
In this letter, we show that URu$_{2}$Si$_{2}$ exhibits clear
signs of quantum criticality upon applying intense magnetic fields
of $\sim$ 40T. Under such extreme conditions, rather than giving
rise to superconductivity, the quantum criticality in
URu$_{2}$Si$_{2}$ gives rise to a plethora of previously-unknown
low-temperature phases, arranged in a hierarchically-organized
phase diagram.

These conclusions follow as the result of a comprehensive
transport study as a function of both temperature (\textit{T}) and
\textit{H} performed in the 45 T hybrid magnet at the National
High Magnetic Field Laboratory in Tallahassee. During the field
sweeps, $H$ is swept mostly at a rate of 2 T/min. This is reduced
to 0.1 T/min between $\mu_{0}H$ =33 and 40 T in order to confirm
the integrity of the $H$-dependent hysteresis. Resistivity
($\rho$) is measured using the standard four-probe method with the
current and $H$ applied along the \textit{c}-axis of an oriented
bar-like URu$_{2}$Si$_{2}$ single crystal, while $T$ is measured
using a precalibrated capacitance sensor.

Figure 1(a) depicts the phase diagram of the heavy fermion system
URu$_{2}$Si$_{2}$ which we propose to be the first such example of
order created at a magnetic-field-tuned QCP. The data in Figs.
1(a) and (b) reveal that URu$_{2}$Si$_{2}$ possesses all the
essential hallmarks of quantum criticality at $T\gtrsim$6 K and
magnetic fields $H$ around 38 T. Most striking, however, is that
this quantum critical region then collapses into a multiplicity of
low temperature phases as the temperature is reduced.

Region I in Fig. 1(a) identifies the upper field limit of the
enigmatic hidden order (HO) phase
\cite{palstra85,broholm87,sugiyama90,ramirez92,amitsuka99,chandra02}.
Specific heat studies \cite{jaime02,kim03} recently revealed the
transition temperature $T_{0}\approx$ 17 K into this phase to be
suppressed with increasing field, terminating at a critical field
$\mu_{0}H_{I}=35.0\pm0.3$ T. This was then shown to be followed by
a new ordered phase, region III, at slightly higher magnetic
fields, which was subsequently found to lie below a metamagnetic
crossover field at higher temperatures\cite{harrison03}, indicated
by solid squares in Fig. 1(a). Region IV was proposed to be a
field-induced recovery of the normal metallic phase, with some or
all of the $f$-electrons aligned by $H$. Each of the
previously-known \cite{jaime02,kim03} phase boundaries into phases
I and III are reproduced in the current study by plotting
extremities in the derivatives of $\partial\rho/\partial H$ and
$\partial\rho/\partial T$ (solid coloured circles in Fig. 1(a)).
Examples of raw $\rho(H)$ and $\rho(T)$ data, from which these
phase boundaries are extracted, are shown in Figs. 2 and 3
respectively. The high sensitivity of $\rho$ to changes in the
ground state of URu$_{2}$Si$_{2}$ enables us to identify another
phase, II, recently detected in ultrasound velocity
measurements\cite{suslov}. Most strikingly, however, our
resistivity measurements also find a complex series of first order
phase transitions (orange lines) and an all-new ordered phase, V,
revealing the hierarchical structure of the phase diagram.
Finally, all phase transitions are found to become hysteretic as a
function of $H$ for $T<$3 K, clearly establishing them to be first
order. Open and solid circles in the inset to Fig. 2 depict
increasing and decreasing $H$ respectively.

It was conjectured on the basis of a magnetization study that
phase III is created in the vicinity of a quantum critical end
point\cite{harrison03}. Although the evidence was limited, if this
holds true, phases II and V could also be the product of quantum
criticality. We will now demonstrate that the present study
reveals evidence for quantum criticality in URu$_{2}$Si$_{2}$ that
is as compelling as reported for any other strongly correlated
metal. We begin by considering the broad resistivity maximum
centered on $\approx$ 34 T at 12 K in Figs. 1(a) and 2. If we
discount the region occupied by the HO phase (below $\approx$ 17 K
and $\approx$ 35 T), this maximum (indicated by + symbols) narrows
and systematically shifts to higher fields as the temperature is
reduced down to $\sim$ 6 K. Its location extrapolates to
$\mu_{0}H\approx37.0\pm0.4$ T in Fig. 1(a) upon fitting a third
order polynomial. Similar resistivity maxima occur in Sr$_{3}$Ru$_{2}$O$_{7}%
$\cite{grigera01} and CeRu$_{2}$Si$_{2}$\cite{kambe96} near their itinerant
metamagnetic transitions; however, they persist down to the lowest temperature
of $\sim$ 0.1 K without any signs of order, in contrast to our URu$_{2}$%
Si$_{2}$ results.

More definitive evidence for quantum criticality is found at
fields greater than 39 T within region IV. While at low
temperatures, region IV can be characterized as a good metal (see
Fig. 3) for which $\rho=\rho_{0}+AT^{n}$ with $n\approx$ 2,
typical of a normal Fermi liquid \cite{kw} (see Fig. 1(b) for low
temperature values of $A$ and $n$), above 6 K the exponent
switches to $n\lesssim$1, which is atypical of a Fermi liquid at
these low temperatures. This crossover gives rise to a broad
maximum in $\partial\rho/\partial T$ which we denote as a
characteristic temperature $T^{\ast}$, delineated by black
triangles in Fig. 1(a). A third order polynomial fit to the field
dependence of $T^{\ast}$ extrapolates to intercept the field axis
at $\mu _{0}H\approx37.1\pm$ 0.5 T inside phase III. Furthermore,
the field-dependence of the Fermi liquid parameter $A$ below
$T^{\ast}$ also appears to be rising to a maximum within phase III
in Fig. 1(b). These observations evidence what is now emerging as
a universal behaviour in correlated metals:\emph{a quantum phase
transition in the Fermi liquid state stabilizes a new ordered
phase, most likely through quantum fluctuations.} The persistent
metallic behaviour within the various ordered phases provides a
unique opportunity for the effect of critical fluctuations on the
quasi-particles to be studied in the presence of ordering. Such a
study cannot be done in systems where the ordered phase is
superconducting, as is the case in heavy fermion metals, tuned by
external pressure, or the high-temperature superconductors, tuned
by doping. Fits of $\rho=\rho_{0}+AT^{n}$ over the temperature
interval between $\sim0.6\leq T\leq$3 K at many different fields
in URu$_{2}$Si$_{2}$, reveal a continuous drop in $n $ from $\sim$
1.5 at 30 T to $\sim$ 1.1 at $\approx$38 T. The normal Fermi
liquid value of $n=2$ is recovered in a discontinuous fashion only
when $\mu_{0}H>$ 39 T.

Exponents that depart significantly from $n=2$ at low temperatures
are generally interpreted as evidence for non-Fermi liquid
behaviour \cite{stewart01}. The continuous trend in $n$ for
$\mu_{0}H<$39 T suggests that the nature of the order parameter in
all of the ordered phases is related. The presence of the
anomalous resistivity exponent $n\sim$1 centered on 38 T further
suggests that, rather than being completely quelled by
ordering, quantum critical fluctuations continue to play a role in URu$_{2}%
$Si$_{2}$ at low temperatures. Whether this finding is linked to the existence
of only a partially gapped Fermi surface, or whether it is generic property of
ordered phases near QCPs, is a pivotal question that remains to be answered.
Such behaviour is reminiscent of the situation in the high-temperature
superconductors, where spin-fluctuation effects persist inside the
superconducting phase \cite{cheong91}.

Collectively, all of the above findings (as summarized in Fig.
1(a)) firmly establish the existence of a field-induced quantum
critical point in URu$_{2} $Si$_{2}$ at $\mu_{0}H=37\pm$1 T. These
findings can be listed as follows: (i) a single metamagnetic
transition observed at temperatures above $\sim$ 6 K at $\sim$37.9
T (solid squares); (ii) a single broad maximum in $\rho(H)$ above
$\sim$6 K (+) that appears to converge with the metamagnetic
transition upon extrapolation to $T=0$; (iii) a crossover
temperature $T^{\ast}$ (from $n\lesssim$ 1 to 2) at fields above
the metamagnetic transition that appears to converge with both the
metamagnetic transition and the magnetoresistance maximum upon
extrapolation to $T=0$ (solid triangles); (iv) a steep increase in
the Fermi liquid parameter $A$ as $H$ approaches phase III from
within region IV; and (v) a gradual reduction of the resistivity
exponent $n$ from $\sim$1.5 to $\sim$1 as $H$ is swept through
successive phases, reaching its lowest value at $\sim$38 T in Fig.
1(b).

By opening an energy gap over nearly the entire Fermi surface,
superconductivity is one of the most efficient means of lowering
the total energy of a metal. However, in spite of the close
similarity of certain generic features of the phase diagram to
those in the cuprates and heavy fermion superconductors, \emph{the
creation of multiple phases near a QCP is presently unique to
URu$_{2}$Si$_{2}$} This may be a simple consequence of the fact
that superconductivity is not favoured because $H$ is difficult
for superconductors to accommodate. What is clear is that the
existence of many rival phases over a small interval of field
indicates many competing ordering mechanisms with similar energy
scales. This eliminates the possibility of there being a single
dominant mechanism; something that may be equally true in the
cuprates as well as the heavy fermion superconductors
\cite{laughlin01}. Our ability to resolve competing mechanisms in
the present study might result from the quantized manner in which
$H$ couples to the various orbital and spin degrees of freedom in
URu$_{2}$Si$_{2}$.

Experiments performed at the NHMFL are supported by the U.S.
National Science Foundation through Cooperative Grant No. DMR
9016241, the State of Florida and the U.S. Department of Energy.
We specially thank B. Brandt for helps in experiments.

\begin{figure}[tbp]
\caption{ (a) High field-phase diagram of URu$_2$Si$_2$ obtained
from our $\rho$ vs. $H$ and $\rho$ vs. $T$ data combined with a
contour plot of the resistivity in which contour values are
indicated. Solid coloured circles (connected by coloured lines)
denote phase transitions extracted from extremities in
$\partial\rho/\partial H$ (from $H$-decreasing sweeps) and
$\partial\rho/\partial T$ ($T$-sweeps); examples of raw data are
shown in Figs. 2 and 3. $+$ symbols indicate the broad maximum in
$\rho(H)$ observed at higher temperatures, solid squares denote
the high-temperature metamagnetic transition field from
magnetization data [20], while solid triangles denote the
crossover temperature, $T^{\ast}$, to $T^2$ behaviour at low $T$
in region IV. Region I refers to the hidden order phase, while II,
III, and V constitute newly discovered phases. (b) A plot of $n$
vs. $H$ (left axis) and $A$ vs. $H$ (right axis) following fits of
the $T$-dependence of $\rho$ for $\sim$ 0.6 K $\leq$ $T$ $\leq$ 3
K in Fig. 3 to the formula $\rho=\rho_{0}+AT^{n}$. The hatched
regions refer to fields where the fitting could not be performed
due to hysteresis. } \label{fig1}
\end{figure}

\begin{figure}[tbp]
\caption{Examples of $\rho$ vs. $H$ close to the QCP in URu$_{2}%
$Si$_{2}$ at selected temperatures and intervals in $H$. Solid and
open circles indicate extremities in $\partial\rho/\partial H$
observed on falling and rising field sweeps, respectively. The
phase boundary lines extracted on rising and falling field sweeps
(and on sweeping $T$) are shown in the inset, evidencing
significant hysteresis in phase V.} \label{fig2}
\end{figure}

\begin{figure}[tbp]
\caption{$\rho$ vs. $T$ data of URu$_{2}$Si$_{2}$ at various
constant magnetic fields, where shift of 10 $\mu\Omega cm$ is made
as indicated for clarity. All data are for increasing temperature
except where the hysteresis is identified at $\mu_{0}H$ = 35.7 T
at $\sim$1.7 K, indicating that hysteresis also occurs as a
function of $T$ on entering or exiting phase V.} \label{fig3}
\end{figure}

\newpage
\clearpage \vspace*{0cm}
\begin{center}
\hspace*{0cm}
\includegraphics[height=22cm,width=17cm,angle=0]{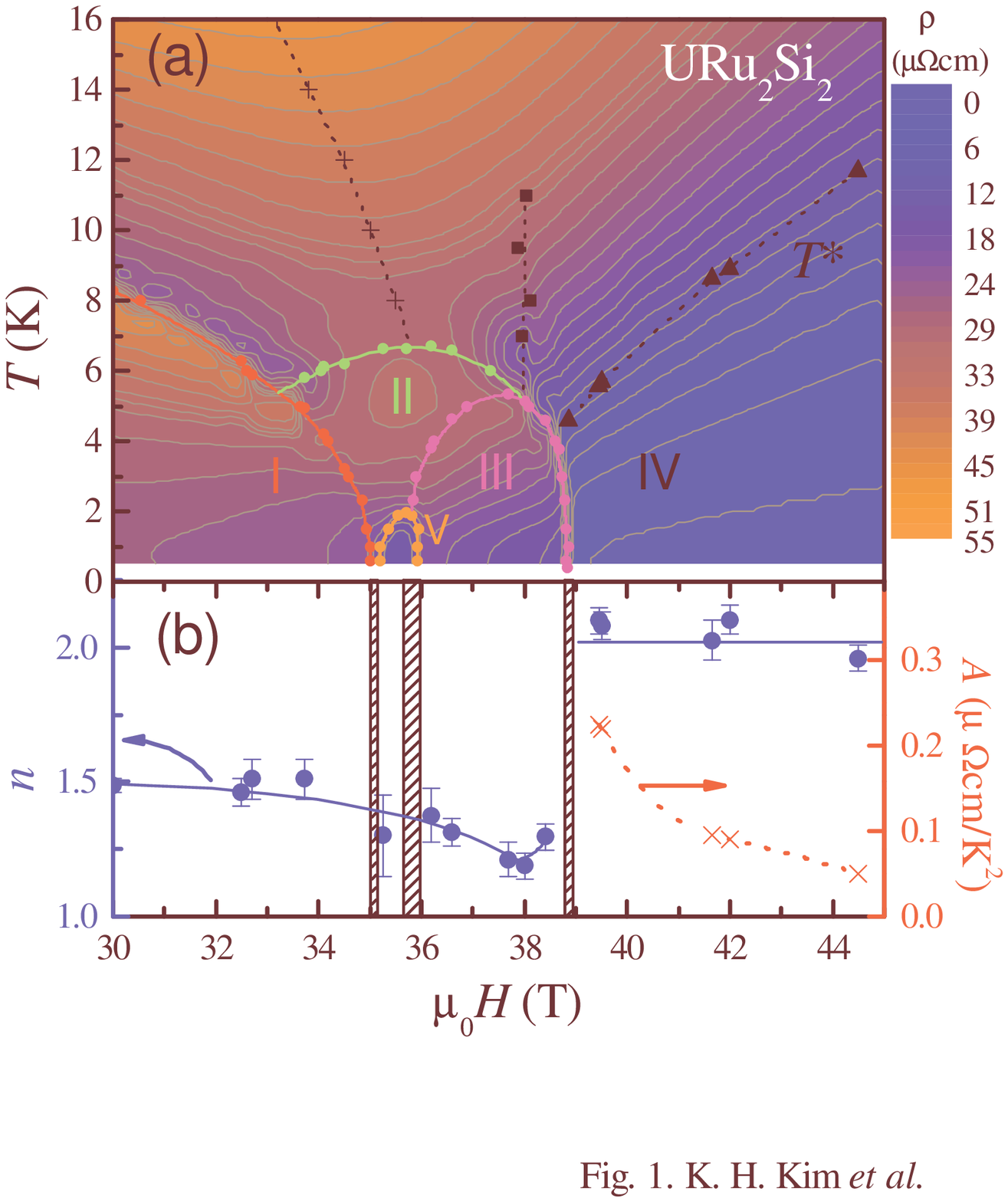}
\newpage
\clearpage \vspace*{0cm}
\end{center}
\begin{center}
\hspace*{0cm}
\includegraphics[height=22cm,width=17cm,angle=0]{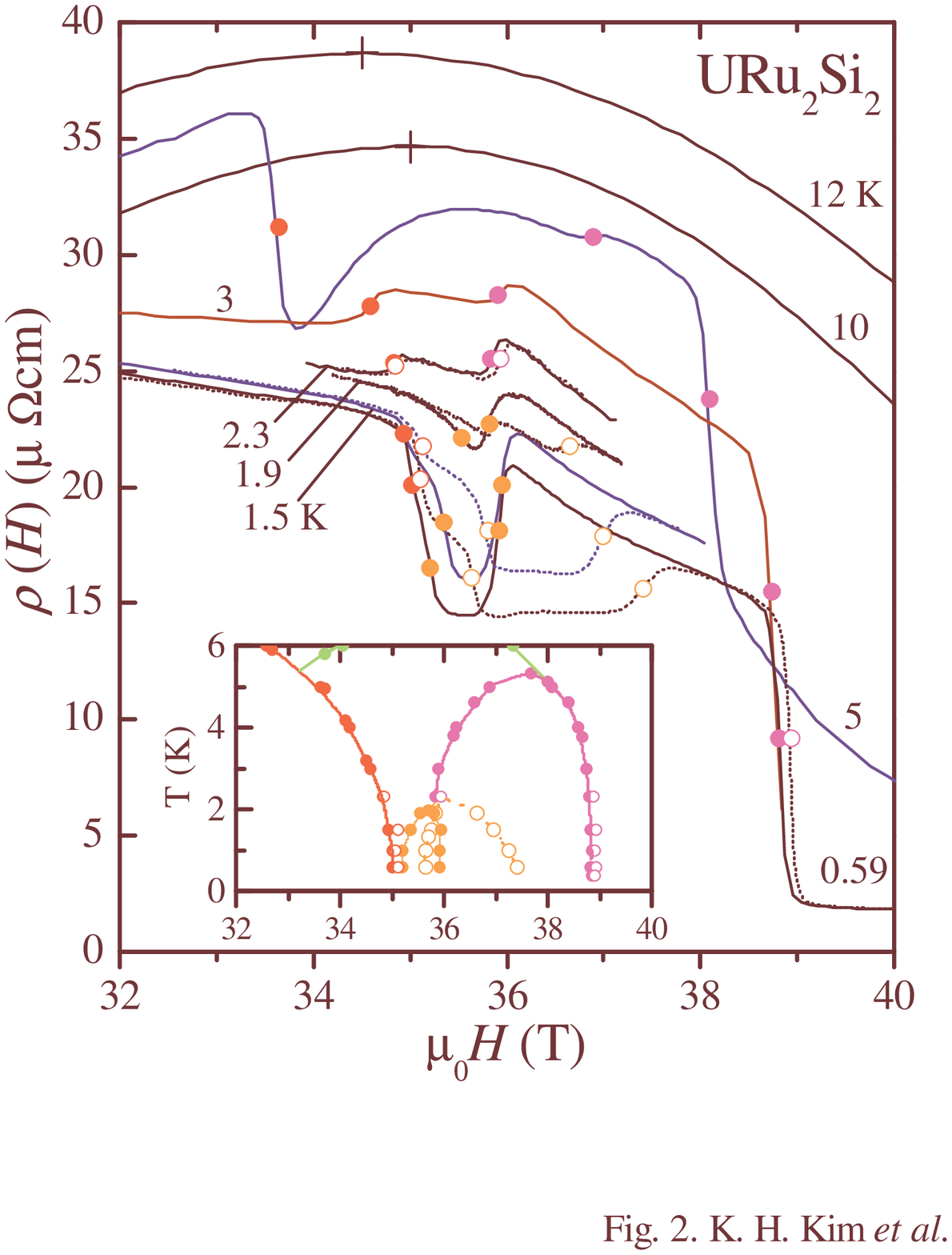}
\end{center}
\newpage
\clearpage \vspace*{0cm}
\begin{center}
\hspace*{0cm}
\includegraphics[height=22cm,width=17cm,angle=0]{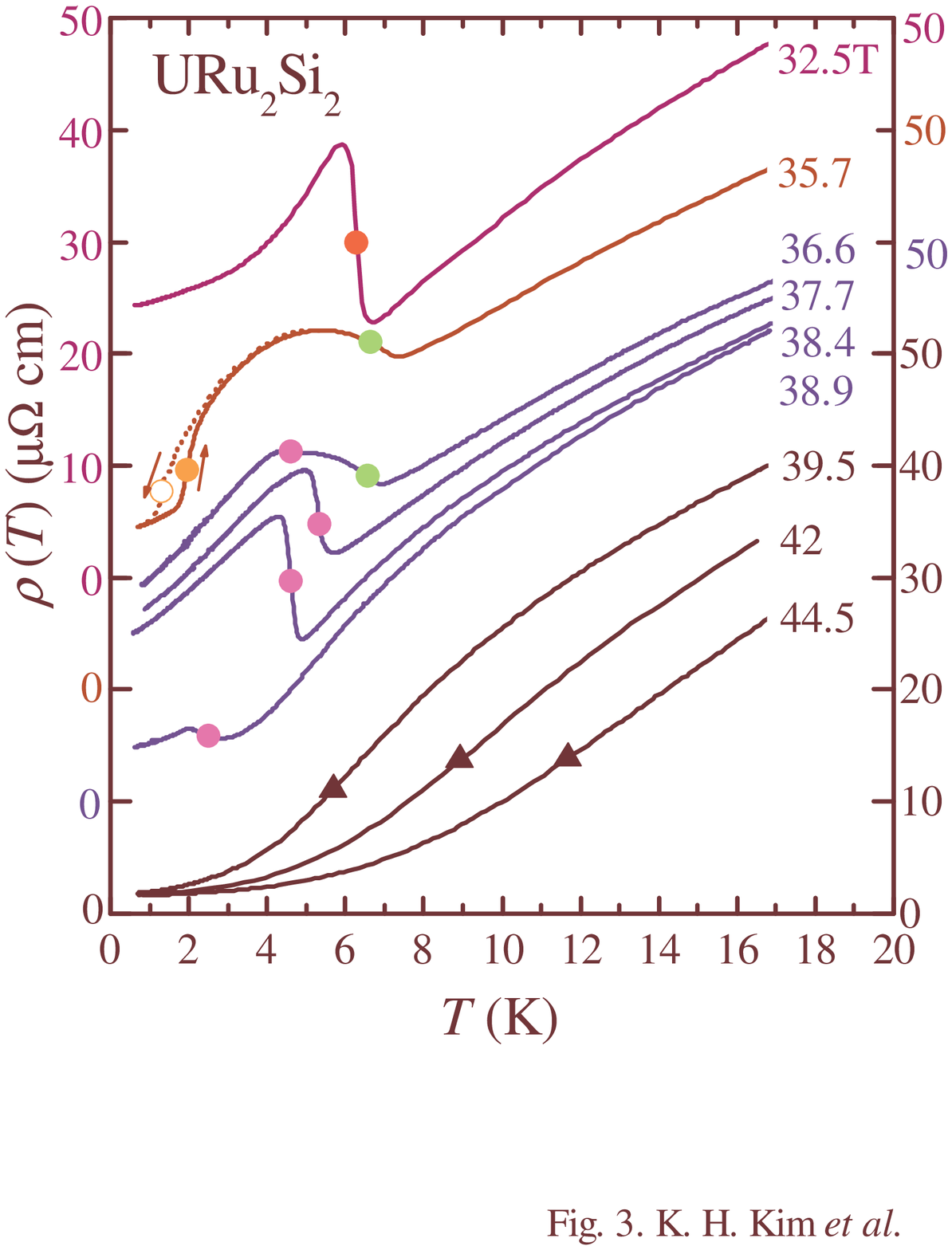}
\end{center}

\begin{references}

\bibitem[*]{khkim} To whom correspondence should be addressed; E-mail: khkim@lanl.gov.
\bibitem {laughlin01}R. B. Laughlin, and D. Pines, Adv. in Physics
\textbf{50}, 361 (2001).

\bibitem {hertz76}J. A. Hertz, Phys. Rev. B \textbf{14}, 1165 (1976).

\bibitem {millis93}A. J. Millis, Phys. Rev. B \textbf{48}, 7183 (1993).

\bibitem {mathur98}N. D. Mathur \textit{et al}., Nature \textbf{394}, 39 (1998).

\bibitem {saxena00}S. S. Saxena \textit{et al}., Nature \textbf{406}, 587 (2000).

\bibitem {lohneysen94}H. v. L\"{o}hneysen \textit{et al}, Phys. Rev. Lett. \textbf{72},
3262 (1994).

\bibitem {sachdev99}S. Sachdev, \emph{Quantum Phase Transitions} (Cambridge
Univ. Press, Cambridge, 1999).

\bibitem {pfleiderer97}C. Pfleiderer, G. J. McMullan, S. R. Julian, and G. G.
Lonzarich, Phys. Rev. B \textbf{55}, 8330 (1997).

\bibitem {tallon01}J. L. Tallon and J. W. Loram, Physica C \textbf{349}, 53 (2001).

\bibitem {kivelson01}S. A. Kivelson, G. Aeppli, and V. J. Emery, Proceed. Nat.
Acad. Sci. \textbf{98}, 11903 (2001).

\bibitem {grigera01}S. A. Grigera \textit{et al}., Science \textbf{294}, 329 (2001).

\bibitem {palstra85}T. T. M. Palstra \textit{et al}., Phys. Rev. Lett. \textbf{55},
2727 (1985).

\bibitem {broholm87}C. Broholm \textit{et al}., Phys. Rev. Lett. \textbf{58},1467 (1987).

\bibitem {sugiyama90}K. Sugiyama \textit{et al}., J. Phys. Soc. Jpn. \textbf{59}, 3331 (1990).

\bibitem {ramirez92}A. P. Ramirez \textit{et al}., Phys. Rev. Lett. \textbf{68}, 2680 (1992).

\bibitem {amitsuka99}H. Amitsuka \textit{et al}., Phys. Rev. Lett. \textbf{83}, 5114 (1999).

\bibitem {chandra02}P. Chandra, P. Coleman, J. A. Mydosh, and V. Tripathi,
Nature \textbf{417}, 831 (2002).

\bibitem {jaime02}M. Jaime, K. H. Kim, G. Jorge, S. McCall, and J. A. Mydosh,
Phys. Rev. Lett. \textbf{89}, 287201 (2002).

\bibitem {kim03}J. S. Kim, D. Hall, P. Kumar, and G. R. Stewart, Phys. Rev. B
\textbf{67}, 014404 (2003).

\bibitem {harrison03}N. Harrison, M. Jaime , and J. A. Mydosh, cond-mat/0301244.

\bibitem {suslov}A. Suslov \textit{et al}., cond-mat/0212158.

\bibitem {kambe96}S. Kambe, H. Suderow, J. Flouquet, P. Haen, and P. Lejay,
Solid State Commun. \textbf{95}, 449 (1995); Corrigendum,
\textit{ibid} \textbf{96}, 175 (1996).

\bibitem {kw}From $A=0.09$ $\mu\Omega$ $cm/K^{2}$ at $H=42$ T and the
Kadowaki-Woods ratio $A/\gamma_{0}^{2}=10^{-5}\mu\Omega$ $cm$ ($mol%
K/mJ)^{2}$ we obtain an expected $\gamma_{0}\approx100$ $mJ/mol$
$K^{2}$ which is in reasonable agreement with the reported
experimental values \cite{jaime02,kim03}.

\bibitem {stewart01}G. R. Stewart, Rev. of Mod. Phys.\textbf{73}, 797 (2001).

\bibitem {cheong91}S.-W. Cheong \textit{et al}., Phys. Rev. Lett. \textbf{67}, 1791 (1991).
\end{references}
\end{document}